\documentclass[12pt]{article}
\usepackage[margin=1in]{geometry} 
\usepackage{amsmath,amsthm,amssymb}
\usepackage{bm}
\usepackage{graphicx}
\usepackage{xcolor}
\usepackage{setspace}
\usepackage{lineno}
\usepackage{cite}
\usepackage{setspace}
\usepackage{hyperref} % 하이퍼링크 기능 활성화

\usepackage[ruled,vlined,linesnumbered]{algorithm2e}
\DontPrintSemicolon  % 줄 끝 ; 생략 (선택)

\hypersetup{
	colorlinks=true,      % 링크에 색상 적용
	linkcolor=blue,       % 내부 문서 링크 색상
	citecolor=blue}

\newcommand{\x}{\mathbf{x}}
\newcommand{\y}{\mathbf{y}}
\newcommand{\n}{\mathbf{n}}
\date{}
\begin{document}
%\doublespacing
%\linenumbers % Activate line numbering
	
% --------------------------------------------------------------
%                         Start here
% --------------------------------------------------------------
 
\title{Sensor placement for sparse force reconstruction}

\author{Jeunghoon Lee\\ 
Dept. of mechanical engineering, Changwon National University \\ jhoonlee@changwon.ac.kr
 %\\ December 15, 2024\\ Last edit: \today
 }

\maketitle

\begin{abstract}
	
	The present study proposes a Gram-matrix-based sensor placement strategy for sparse force reconstruction in the frequency domain. A modal decomposition of the Gram matrix reveals that its structure is dominated by a few modes near the target frequency, and that each modal contribution reflects the spatial correlation of the corresponding mode shape. This suggests that placing sensors near nodal regions where spatial correlation is low can reduce coherence in the frequency response function (FRF) matrix and improve force reconstruction accuracy. To translate the physical insight into a practical design framework, a greedy algorithm is proposed to  select sensor locations that minimize the off-diagonal energy of the Gram matrix. Numerical simulations and experimental validations demonstrate that the proposed method yields robust and accurate force estimation,  outperforming heuristic sensor layouts.

\end{abstract}


\vspace{1cm}
\section*{Highlights}
\begin{itemize}
	\item Sensor placement using Gram matrix is proposed for force reconstruction.
	\item Sensors are placed at nodal points of mode-shape near the target frequency.
	\item The method reduces measurement correlation and enhances reconstruction accuracy.
\end{itemize}

\vspace{0.5cm}

\section{Introduction} \label{sec:intro}

Accurate force reconstruction is a critical issue in structural health monitoring, vibration control, and many engineering applications. The location of the applied force is generally unknown, and even when known, direct force measurement is infeasible for most structures. Alternatively, indirect estimation using limited sensor data (e.g., accelerometers) and a physical model leads to an inverse problem, where the number of unknown forces to be estimated often exceeds the available sensor measurements~\cite{sanchez2014}. Sparse signal processing, known as the compressed sensing (CS), addresses such underdetermined inverse problems by assuming that the solution is sparse, i.e., only a few non-zero elements exist~\cite{donoho2006}. The sparsity assumption is valid for many source identification problems, as sources typically occupy a small spatial extent within the entire domain. Thus, CS has been successfully applied to source identification in the fields of vibration and acoustics~\cite{gerstoft2018introduction,xenaki2014compressive,lee2023cfd, antoni2012}. 

CS-based force identification—hereafter termed sparse force reconstruction—has been developed in two major forms: time-domain~\cite{qiao2016,qiao2017,qiao2019,liu2023} and frequency-domain~\cite{zhang2012,aucejo2014,aucejo2016,aucejo2017} approaches. The former is suitable for reconstructing transient or impact forces, whereas the latter are designed for steady-state or periodic excitations. The present study focuses on frequency-domain reconstruction of spatially sparse forces, i.e., point excitations acting at a small number of discrete locations on the structure. Prior works have primarily focused on solving the inverse problem and improving algorithmic stability and efficiency.

%CS-based force identification—termed sparse force reconstruction—can be classified into time-domain~\cite{qiao2016,qiao2017,qiao2019,liu2023} and frequency-domain~\cite{zhang2012,aucejo2014,aucejo2016,aucejo2017} approaches. The former is suitable for reconstructing transient or impact forces, while the latter focuses on steady-state or periodic loads. This study emphasizes frequency-domain reconstruction, for which numerous formulations have been proposed depending on load characteristics (e.g., point or distributed). Prior works have primarily focused on solving the inverse problem and improving algorithmic stability and efficiency.

%Yet, sensor placement for accurate reconstruction has received little attention. While CS allows for fewer sensors, reconstruction performance is sensitive to their placement. In the CS literature, particularly in acoustic source localization, sensor placement strategies have been proposed to minimize the restricted isometry property (RIP)~\cite{lee2019optimalsensor}, asserting that sparse signals are mapped with minimal distortion. As well, various sensor configurations have been studied for direction-of-arrival (DOA) estimation~\cite{xenaki2014compressive}. This is not the case for force reconstruction problems, where systematic investigations into sensor placement are still lacking.
Yet, sensor placement for accurate reconstruction has received little attention. While CS allows for fewer sensors, reconstruction performance is sensitive to their placement. In the CS literature, particularly in acoustic source localization, sensor placement strategies have been proposed to minimize the restricted isometry property (RIP)~\cite{lee2019optimalsensor}, asserting that sparse signals are mapped with minimal distortion. Several sensor configurations have also been studied for direction-of-arrival (DOA) estimation~\cite{xenaki2014compressive}, where the concern is on exploring different array geometries under homogeneous propagation conditions. By contrast, force reconstruction in mechanical systems is governed by structural modal characteristics: the FRF basis functions depend on eigenmodes, and sensor placement is linked to modal overlap and nodal distributions. Systematic investigations into sensor placement are still lacking

Conventional practices involve deploying as many sensors as possible to maximize observability, a heuristic that is often impractical. Noteworthy efforts include the Effective Independence (EI)-based methods~\cite{kammer1991,stephan2012,kim2024}, designed for sensor placement in modal testing. These approaches utilize the Fisher Information Matrix (FIM) to measure the sensitivity of the system response with respect to parameter variations. The EI index derived from the FIM identifies sensors that are most informative for representing the system’s dynamic behavior. The strategy favors sensor placement at anti-nodal points of dominant mode shapes, where response amplitudes are high—making it well-suited for forward problems such as response estimation. As will be demonstrated, both the dense deployment and the anti-nodal scheme yield poorer reconstruction than the coherence-driven approach proposed here.

A key requirement for sparse recovery is the incoherence of the sensing matrix, which ensures that the basis vectors (i.e., columns of the sensing matrix) are as uncorrelated as possible~\cite{foucart2013}. In the context of force reconstruction, the sensing matrix is given by the frequency response function (FRF) matrix. High coherence among basis vectors can lead to ambiguous outputs and degraded reconstruction accuracy.

The present study proposes a Gram-matrix-based strategy~\cite[Chap.~5]{foucart2013}, where the Gram matrix quantifies coherence among basis vectors of the FRF matrix. A modal decomposition of the Gram matrix reveals that its structure is dominated by a few modes near the given target frequency, each contributing through the spatial correlation of mode shape. This suggests that placing sensors near nodal points promotes independence among measurements. For systematic sensor placement, a greedy algorithm is introduced to minimize the off-diagonal elements of the Gram matrix, guiding sensor placement toward the nodal regions of dominant modes associated with the target frequency. Tailored for inverse force reconstruction, the proposed method offers a unique alternative to conventional forward-oriented sensor placement strategies. To the best of the author's knowledge, it is the first comprehensive approach that links Gram-matrix analysis to sensor placement for sparse force reconstruction.

The remainder is organized as follows. Section~\ref{sec:theory} formulates the sparse force reconstruction problem within the CS framework. Section~\ref{sec:gram} presents a modal analysis of the Gram matrix, offering physical insight into sensor placement. Section~\ref{sec:greedy} introduces a greedy optimization algorithm for selecting sensor locations. Sections~\ref{sec:numerical} and~\ref{sec:exp} provide numerical and experimental validations, respectively.

\section{Sparse force reconstruction} \label{sec:theory}
\subsection{System model}
For a damped $N$ dof (degree-of-freedom) system, the frequency-domain relationship between the force excitation at $n$-th node and acceleration response at $m$-th node is described by the following FRF \cite[Chap.~2]{ewins2000}:

\begin{equation} \label{eq:scalarFRF}
	h_{mn}(\omega)=\sum_{r=1}^{N}\frac{-\omega^2A_{mn}^{(r)}}{\omega_r^2-\omega^2+j2\zeta_r\omega_r\omega},
\end{equation}
where $j=\sqrt{-1}$, and $\omega =2\pi f$ denotes the angular frequency (with $f$ in Hertz). FRF comprises the contributions of $N$ vibration modes indexed by $r$, with each mode characterized by its undamped natural frequency $\omega_r$, damping ratio $\zeta_r$, and modal constant $A_{mn}^{(r)}$. In general, $A_{mn}^{(r)}$ is a complex-valued quantity that describes the coupling strength between the input $(n)$ - output $(m)$ pair for the $r$-th mode and determines how the $r$-th mode contributes to the response. Those modal parameters $(\omega_r, \zeta_r, \text{and} \,  A_{mn}^{(r)})$ are extracted through the modal analysis~\cite[Chap.~3,4]{ewins2000} performed on either a numerical or experimental model.

A useful indicator for quantifying the spectral overlap between neighboring modes is the modal overlap factor (MOF)~\cite{mace2005,ege2009}, defined as
\begin{equation} \label{eq:MOF}
	\mathrm{MOF} = \frac{2\zeta_r \omega}{\omega_{r+1} - \omega_r}.
\end{equation}
A low MOF indicates that the modes are spectrally well separated, whereas a high MOF suggests significant modal interaction. This factor will later play an important role in assessing the validity of modal approximations, as well as the stability in force reconstruction.

We assume a system with proportional damping, where the damping matrix is represented by a linear combination of the system's mass and stiffness matrices. It preserves the useful orthogonal property of undamped systems. Specifically, the mode-shape matrix satisfies the following orthonormality condition relative to the mass matrix $\mathbf{M} \in \mathbb{R} ^ {N \times N}$:
\begin{equation} \label{eq:ortho_property}
	\mathbf{\Phi}^\mathsf{T} \mathbf{M}\mathbf{\Phi} = \mathbf{I}_N, 
\end{equation}
where $^\mathsf{T}$ is the transpose, $\mathbf{\Phi} \in \mathbb{R} ^ {N \times N}$ is the matrix of mode-shapes, and $\mathbf{I}_N$ is $N \times N$ identity matrix. Moreover, the modal constant $A_{mn}^{(r)}$ becomes a real-valued quantity, given by the product of two terms:
\begin{equation} \label{eq:Amn}
	A_{mn}^{(r)} = \phi_{mr}\phi_{nr},
\end{equation}
where the mode-shape coefficient $\phi_{ij}$ denotes the $i$-th row and $j$-th column element of $\mathbf{\Phi}$.

The proportional damping model is adopted to simplify the modal decomposition of the Gram matrix in Section~\ref{sec:gram_decompose}. It should be emphasized, however, that the proposed sensor placement strategy is not limited to this simplified case. To demonstrate the method’s applicability to more general  scenarios, a numerical example involving non-proportional damping is presented in Section~\ref{sec:sim_irregular}.

\subsection{Force reconstruction via LASSO}
Force reconstruction refers to the determination of $N$ unknown forces using a few acceleration measurements $M (< N)$. Frequency-domain representation for the measurement vector $\y \in \mathbb{C}^M$ is as follows:
\begin{equation} \label{eq:linear_model}
	\y(\omega) = \mathbf{H}(\omega) \x(\omega) + \n(\omega),
\end{equation}
where the vector $\x \in \mathbb{C}^N$ that we aim to reconstruct contains unknown forces for all $N$ nodes. 
The noise vector  $\n \in \mathbb{C}^M$, assumed to be  independent across the sensors, encompasses various sources of error, including measurement noise from experiments and inaccuracies inherent in numerical models. Using Eq.~\eqref{eq:scalarFRF}, the FRF matrix $\mathbf{H} \in \mathbb{C}^ {M \times N}$ (termed the \textit{sensing} matrix in CS community) is expressed by
\begin{equation}
	\mathbf{H}=[\mathbf{h}_1 \, \mathbf{h}_2 \, \ldots \, \mathbf{h}_N],
\end{equation}
where the vector $\mathbf{h}_n = [h_{1n} \, h_{2n} \, \ldots \, h_{Mn}]^{\mathsf{T}} \in \mathbb{C}^{M}$ (termed the \textit{basis} vector hereinafter) collects frequency responses at all measurement points for $n$-th input node. 

We adopt the basic framework for CS-based sparse force reconstruction, known as the least absolute shrinkage and selection operator (LASSO)~\cite{tibshirani1996}. The technique solves a convex least-square problem augmented by sparsity-promoting $\ell_1$-norm regularizer. Using the linear model in Eq.~\eqref{eq:linear_model}, LASSO is formulated as:
\begin{equation} \label{eq:lasso}
	\hat{\x} = \underset{\x \in \mathbb{C}^N}{\arg \min} \left[J= \frac{1}{2} \Vert\mathbf{Hx} - \y\Vert_2^2 + \mu \Vert\x\Vert_1\right],
\end{equation}
where $\Vert\x\Vert_p=({\sum\nolimits_{n = 1}^N |x_n|^p})^{1/p}$ denotes the $\ell_p$-norm. The regularization parameter $\mu$ balances the trade-off between solution sparsity and data fidelity; a high $\mu$ promotes the solution sparsity with sacrificing the data-fidelity, and vice-versa. From a Bayesian perspective, the solution $\hat{\x}$ represents the maximum a posteriori (MAP) estimate under Laplace prior~\cite[Sec.~11.4]{pml1Book}, reflecting that most of the solution coefficients are zero. Different priors can be adopted according to the solution characteristics, resulting in different formulations beyond LASSO~\cite{aucejo2016,xenaki2016block}. 

The measurements are typically small quantity, which hinders the selection of an appropriate $\mu$ and causes a numerical instability in solving the optimization problem. To remove the amplitude dependence, the cost $J$ is divided by the squared $\ell_2$-norm of the measurement vector $\Vert \y\Vert_2^2$. Also note that the basis vectors have different $\ell_2$-norms, which can introduce bias into the reconstructed solution~\cite[Sec.~3.1.4]{elad2010}. In the context of the force reconstruction, this bias may causes an overestimation of forces located near the sensors, while forces farther away are likely to be underestimated. Thus, the basis vectors are normalized to have unit $\ell_2$-norm by performing the operation $\overline{\mathbf{H}}=\mathbf{HF}$, where $\mathbf{F}$ is a $N \times N$ diagonal matrix containing $1/{\Vert \mathbf{h}_n\Vert_2}$ as its diagonal entries, i.e., $\mathbf{F}= \text{diag} \big[ 1/{\Vert \mathbf{h}_1\Vert_2}, \,\, 1/{\Vert \mathbf{h}_2\Vert_2}, \,\, \ldots, \,\, 1/{\Vert \mathbf{h}_N\Vert_2}\big]$. Applying this column normalization and scaling the measurement modify the original LASSO problem in Eq.~\eqref{eq:lasso} into the following generalized LASSO problem~\cite{tibshirani2011genlasso,lee2024pd}:
\begin{equation} \label{eq:lasso_normal}
	\hat{\overline{\x}} = \underset{\overline{\x} \in \mathbb{C}^N}{\arg \min} \left[\overline{J}= \frac{1}{2} \Vert\mathbf{\overline{H}\overline{x}} - \overline{\y}\Vert_2^2 + \overline{\mu} \Vert\mathbf{F}\overline{\x}\Vert_1\right],
\end{equation}
where 
\begin{equation}
	\overline{J} = J/\Vert \y \Vert_2^2, \, \overline{\y} = \y/\Vert \y \Vert_2, \, \overline{\mu}=\mu/\Vert \y \Vert_2, \, \overline{\x} = \mathbf{F}^{-1}\x/\Vert \y \Vert_2,
\end{equation}
 and $\mathbf{F}^{-1}= \text{diag} [\{\Vert\mathbf{h}_n\Vert_2\}_{n=1}^N]$. 
 The above is formulated for a single snapshot at a specific frequency. When multiple snapshots across various frequencies are available, $\hat{\overline{\x}}$ is estimated sequentially for each frequency and snapshot. Once $\hat{\overline{\x}}$ is found, the original solution can be recovered through a de-normalization step using $\hat{\x}=\Vert \y \Vert_2 \mathbf{F} \hat{\overline{\x}}$. 
 
 Note that $\overline{J}$ is a convex function, so Eq.~\eqref{eq:lasso_normal} is a convex optimization problem.  Several open-source codes are available for solving such problems~\cite{daubechies2004,figueiredo2007,boyd2011admm,cvx}. This work uses the widely adopted CVX toolbox~\cite{cvx}, implementing the
 interior-point method.

\section{Gram matrix analysis} \label{sec:gram}

\subsection{Definition and modal decomposition} \label{sec:gram_decompose}
For a stable reconstruction, the coherence among the basis vectors (the columns of $\overline{\mathbf{H}}$) is of fundamental importance~\cite{xenaki2014compressive,lee2023cfd,lee2019optimalsensor,gerstoft2015multiple}. If the basis vectors are  coherent to each other, the estimated coefficient $\hat{\x}$ spreads across the indices corresponding to these vectors. This eventually generate spurious estimates at locations where no actual forces exist. Evaluating the coherence of basis vectors is essential for not only achieving accurate reconstruction, but also optimizing sensor placement. As a measure of coherence between any two basis vectors, we use the Gram matrix~\cite{foucart2013,xenaki2014compressive}, 
\begin{equation} \label{eq:gram_definition}
	\mathbf{G}(\omega)=\overline{\mathbf{H}}^\mathsf{H}(\omega)\overline{\mathbf{H}}(\omega) \in \mathbb{C}^{N \times N},
\end{equation}
where $^\mathsf{H}$ denotes the Hermitian transpose. The element of $\mathbf{G}$, $G_{ij}$, represents the inner product between the $i$-th and $j$-th column of $\overline{\mathbf{H}}$, i.e., $G_{ij}=\overline{\mathbf{h}}_i^\mathsf{H}\overline{\mathbf{h}}_j$. Since the basis vectors are normalized to have unit $\ell_2$-norm, $G_{ij}$ represents the cosine of the angle between them. Thus, $\vert G_{ij} \vert$ ranges from 0 to 1, where 0 indicates orthogonality for the considered basis vectors, and 1 indicates complete correlation. The ideal scenario is achieved when $\mathbf{G} = \mathbf{I}_N$. 

Sensor placement involves selecting $M$ nodes from $N$ candidates, with the goal of maintaining independence among the basis vectors of the sensing matrix. Since $M < N$, complete independence is unattainable; in other words, $\mathbf{G} $ cannot be an identity matrix and will inevitably exhibit nonzero off-diagonal elements. A practical strategy for sensor placement is therefore to select $M$ rows from the full sensing matrix $\overline{\mathbf{H}}_{full} \in \mathbb{C}^{N \times N}$ such that the off-diagonal elements of the reduced Gram matrix are minimized.

The element of $\mathbf{G}$ can be analyzed using Eq.~\eqref{eq:scalarFRF}:
\begin{subequations} \label{eq:gram_expansion}
	\begin{align}
		G_{ij}&= \frac{\omega^4}{\Vert \mathbf{h}_i\Vert_2 \Vert \mathbf{h}_j\Vert_2}  \sum_{r=1}^{M}\sum_{p=1}^{N}\sum_{q=1}^{N}\frac{A_{ri}^{(p)}}{\omega_p^2-\omega^2+j2\zeta_p\omega_p\omega} \cdot \frac{A_{rj}^{(q)}}{\omega_q^2-\omega^2+j2\zeta_q\omega_q\omega} \\ 
		&=\frac{\omega^4}{\Vert \mathbf{h}_i\Vert_2 \Vert \mathbf{h}_j\Vert_2}\Bigg[\underbrace{\sum_{r=1}^{M} \sum_{p=1}^{N} \frac{A_{ri}^{(p)}  A_{rj}^{(p)}}{(\omega_p^2 - \omega^2+j2\zeta_p\omega_p\omega)^2}}_{\text{Self-term}} 
		+ \label{eq:gram_expansion2} \\ \nonumber
		& \hspace{3cm}\underbrace{\sum_{r=1}^{M} \sum_{p=1}^{N} \sum_{\substack{q=1 \\ q \neq p}}^{N} \frac{A_{ri}^{(p)}}{\omega_p^2 - \omega^2+j2\zeta_p\omega_p\omega} \cdot \frac{A_{rj}^{(q)}}{\omega_q^2 - \omega^2+j2\zeta_q\omega_q\omega}}_{\text{Cross-term}}\Bigg] \\ 
		%	&\cong \frac{1}{\Vert \mathbf{h}_i\Vert_2 \Vert \mathbf{h}_j\Vert_2}\sum_{r=1}^{M} \sum_{p=1}^{N} \frac{A_{ri}^{(p)}  A_{rj}^{(p)}}{(\omega_p^2 - \omega^2)^2+j2\zeta_p\omega_p\omega}. 	\\
		&\cong \frac{\omega^4}{\Vert \mathbf{h}_i\Vert_2 \Vert \mathbf{h}_j\Vert_2}\sum_{r=1}^{M} \sum_{p=1}^{N} \frac{A_{ij}^{(p)}\phi_{rp}^2}{(\omega_p^2 - \omega^2)^2+j2\zeta_p\omega_p\omega}. \label{eq:gram_expansion3}					
	\end{align}
\end{subequations}

Eq.~\eqref{eq:gram_expansion2} separates the summation into two parts: the self-term and the cross-term. The self-term contains contributions from identical modes, while the cross-term reflects the interactions between different modes. The cross-terms are negligible only when both the modal overlap factor (MOF) is low (e.q., $\mathrm{MOF} < 0.5$ as a conventional criterion) and the system is away from resonance conditions. At resonance frequencies, the denominator $(\omega_p^2 - \omega^2 + j 2\zeta_p \omega_p \omega)$ becomes small, which amplifies both the self- and cross-terms, making the latter non-negligible. The approximation in Eq.~\eqref{eq:gram_expansion3} is obtained by neglecting the cross-terms and applying the relation $A_{ri}^{(p)}A_{rj}^{(p)} = A_{ij}^{(p)}\phi_{rp}^2$ (valid under the proportional damping, Eq.~\eqref{eq:Amn}).

A full measurement scenario ($M = N$) enables the complete capture of all modes without any information loss. Using the orthogonality condition $\sum_{r=1}^{M(=N)} \phi_{rp}^2 = 1 / M_{pp}$, where $M_{pp}$ denotes the $p$-th diagonal element of the mass matrix $\mathbf{M}$ (Eq.~\eqref{eq:ortho_property}), Eq.~\eqref{eq:gram_expansion3} can be expressed as:

\begin{equation} \label{eq:simpleGij}
	G_{ij}\big|_{M=N} \cong \frac{\omega^4}{\Vert \mathbf{h}_i\Vert_2 \Vert \mathbf{h}_j\Vert_2}\sum_{p=1}^N\frac{A_{ij}^{(p)}}{M_{pp}(\omega_p^2 - \omega^2 + j 2\zeta_q \omega_q \omega)^2}.
\end{equation}

This leads to a compact representation of the full Gram matrix $\mathbf{G}_{full}$:
\begin{subequations} \label{eq:simpleG}
	\begin{align}
		\mathbf{G}_{full} &= \overline{\mathbf{H}}_{full}^\mathsf{H} \, \overline{\mathbf{H}}_{full} \label{eq:simpleGa}\\
		&\cong \omega^4 \, \mathbf{F}^\mathsf{T} \left[ \sum_{p=1}^N \frac{\boldsymbol{\phi}_p \boldsymbol{\phi}_p^\mathsf{T}}{M_{pp}(\omega_p^2 - \omega^2 + j 2\zeta_p \omega_p \omega)^2} \right] \mathbf{F}, \label{eq:simpleGb}
	\end{align}
\end{subequations}
where $\boldsymbol{\phi}_p$ denotes the $p$-th column of the mode-shape matrix $\mathbf{\Phi}$. As mentioned, the decomposition in Eq.~\eqref{eq:simpleGb} assumes negligible contribution from cross-terms, a condition satisfied under low MOF and at frequencies away from resonance.

While it lacks generality, the composition in Eq.~\eqref{eq:simpleGb} provides valuable insight into the structure of the Gram matrix. First, for a given frequency $\omega$, $\mathbf{G}_{full}$ is mainly influenced by a few modes whose natural frequencies are near $\omega$. Modes distant from $\omega$ contribute less due to the frequency-dependent attenuation in the denominator. Second, more importantly, each mode contributes via the term $\boldsymbol\phi_p \boldsymbol\phi_p^\mathsf{T}$, representing the spatial correlation of $p$-th mode-shape. This symmetric matrix, i.e., the auto-correlation matrix of $\boldsymbol\phi_p$, serves as a fundamental building block for $\mathbf{G}_{full}$. 

As $\boldsymbol\phi_p \boldsymbol\phi_p^\mathsf{T}$ encodes the outer product of the mode shape with itself, $G_{ij}$ is proportional to $\phi_{ip}\phi_{jp}$, which vanishes or becomes small when either $\phi_{ip}$ or $\phi_{jp}$ approaches zero. That is, the off-diagonal entries of the Gram matrix are suppressed when both measurement points $i$ and $j$ are located near the nodal regions. Given that the goal of sensor placement is to minimize the off-diagonal energy of the reduced Gram matrix, these observations suggest that placing sensors near the nodal regions of dominant modes is advantageous. Such placement reduces spatial correlation between basis vectors, thereby improving incoherence and promoting robust sparse recovery.

In forward problems such as mode-shape identification, it is desirable to place response sensors at anti-nodal points, as the high amplitude provides a high signal-to-noise ratio~\cite{kammer1991,stephan2012,kim2024}. The situation is different in inverse problems whose objective is to reconstruct the input force from measured responses. Placing sensors at anti-nodal locations introduces strong correlations among sensor outputs due to the spatially coherent nature of mode shapes. High correlations lead to poor conditioning of the sensing (FRF) matrix and may generate ghost sources. Conversely, measurements taken at or near nodal points tend to be less affected by mode shapes and exhibit lower inter-sensor correlation. While the absolute response level at nodal points is smaller, these locations are more sensitive to localized force variations. Weak spatial correlation is advantageous for isolating force contributions and improving the robustness of reconstruction.

\begin{figure}[t]
	\centering
	\includegraphics[width=0.95\linewidth]{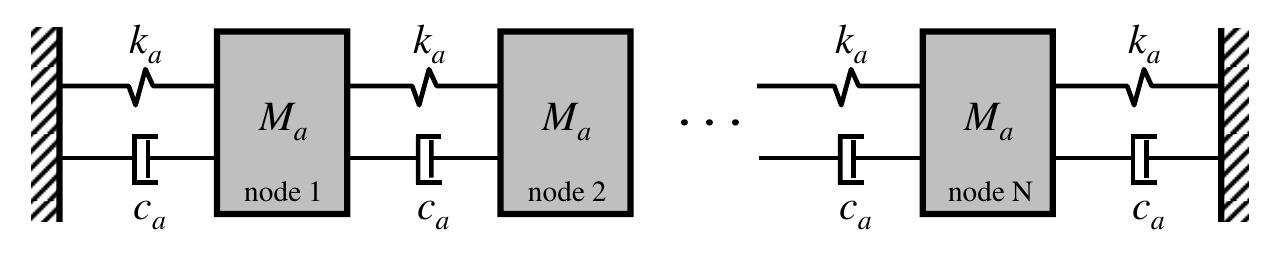}
	\caption{$N$-dof mass - spring - dashpot system. $N=50$, $M_a$ = 2 [kg], $k_a=2 \times 10^6$ [N/m], $c_a= \alpha M_a + \beta k_a$ [N$\cdot$s/m] ($\alpha= 1 \times 10^{-4}$ and $\beta=  1 \times 10^{-3}$)}
	\label{fig:spatial_model}
\end{figure}

\subsection{Example} \label{subsec:GramExample}
For illustration, let us consider a simple regular structure of $N=50$ dof system with fixed ends (Figure~\ref{fig:spatial_model}). Each mass, $M_a$ = 2 kg, is connected in series by springs with a stiffness $k_a=2 \times 10^6$ N/m. Viscous dashpots, placed in parallel to the springs, have a damping coefficient $c_a= \alpha M_a + \beta k_a$ N$\cdot$s/m, where $\alpha = 1 \times 10^{-4}$ and $\beta = 1 \times 10^{-3}$. This configuration leads to a lightly damped system, with the modal damping ratios  ($\zeta_r$) increasing from 0.01 to 0.07. Figure~\ref{fig:overall}(a) shows a sample driving-point FRF (evaluated at node 12), along with the MOF.

\begin{figure}[h]
	\centering
	\includegraphics[width=0.9\linewidth]{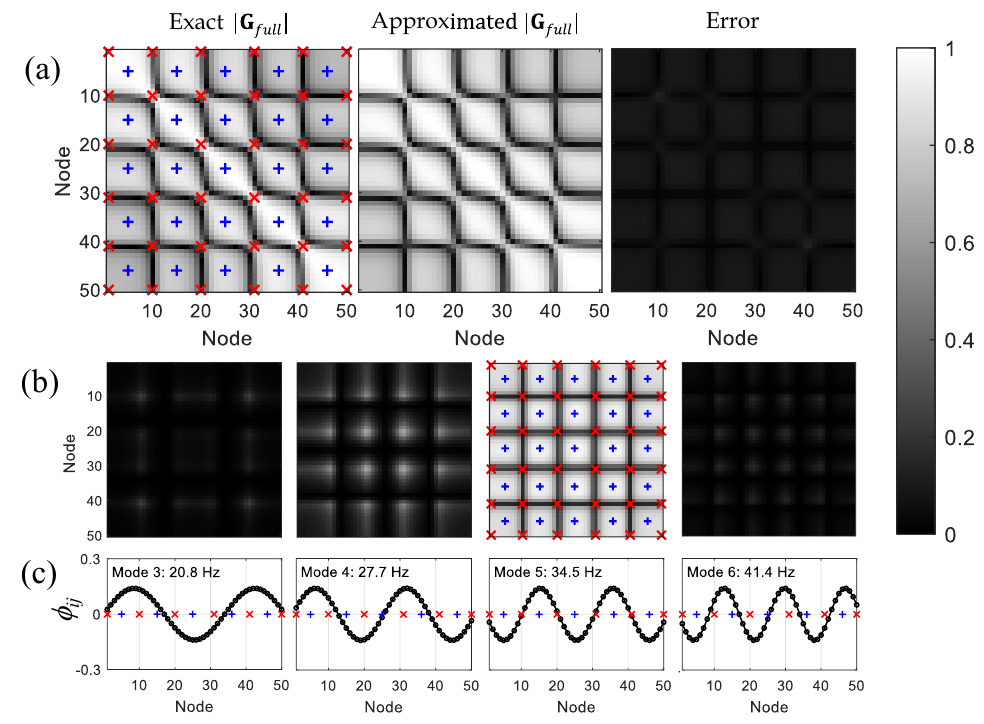}
	\caption{ 
		(a) Left : Full Gram matrix $\big| \mathbf{G}_{full}\big|$ calculated using Eq.~\eqref{eq:gram_definition} for the system shown in Figure~\ref{fig:spatial_model} ($f= 0.95 \omega_5/(2\pi)= 32.8 \text{ Hz} $), 		
			Middle: Approximation of $\mathbf{G}_{full}$ by summing the Gram matrices of four nearby modes (Modes 3–6).  
			Right: Element-wise error between the exact and approximated Gram matrices.  			
		 	(b) Individual contributions from these four modes, with associated mode-shape vectors shown below in (d).
			The marker \textcolor{red}{$\boldsymbol{\times}$} indicates the nodal locations of the fifth mode, while \textcolor{blue}{$\boldsymbol+$} denotes the anti-nodal points.		 	
		}		 
	\label{fig:gramDemo}
\end{figure}
Assuming a full measurement, Figure~\ref{fig:gramDemo}(a, left) shows the Gram matrix $\mathbf{G}_{full}$ computed at $f = 32.8 (= 0.95 \omega_5 / 2\pi)$ Hz. Figure~\ref{fig:gramDemo}(a, middle) displays the Gram matrix obtained by summing the individual contributions from four nearby modes (Modes 3--6). These individual Gram matrices are shown in Figure~\ref{fig:gramDemo}(b), and their respective mode shapes are plotted in Figure~\ref{fig:gramDemo}(c). As shown in Figure~\ref{fig:gramDemo}(a, right), the element-wise approximation error is small. The close agreement validates the modal decomposition in Eq.~\eqref{eq:simpleGb}, which is attributed to the low MOF ($\cong 0.1$) and the avoidance of resonance.
As indicated by the \textcolor{red}{$\boldsymbol{\times}$} markers, nodal points of the fifth mode coincide with regions where the off-diagonal entries of Gram matrix are suppressed, whereas the  \textcolor{blue}{$\boldsymbol+$} markers at anti-nodal points correspond to regions of high coherence. This copes with the physical insight gained from the modal decomposition.

%^(as indicated by \textcolor{blue}{$\boldsymbol+$})

\begin{figure}[t]
	\centering
	\includegraphics[width=0.8\linewidth]{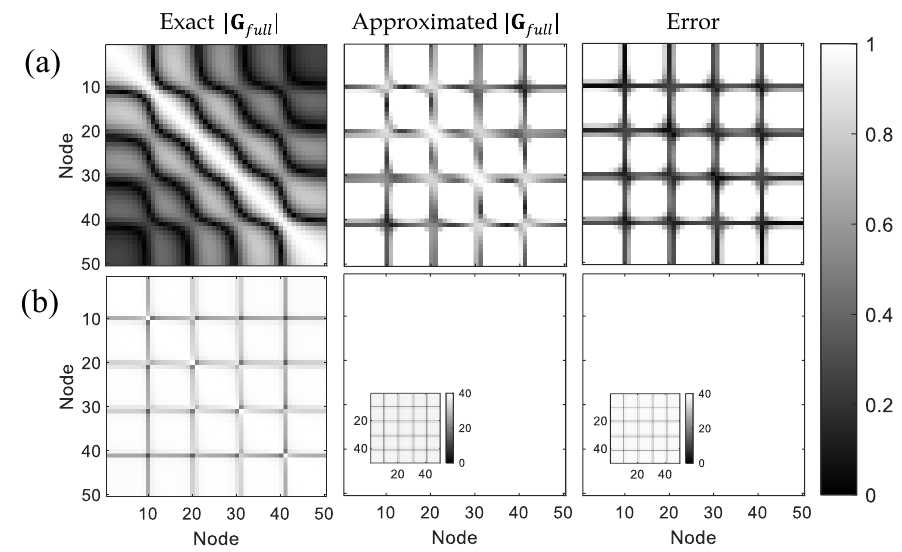}
	\caption{
		Effects of high MOF and resonance on the Gram matrix structure and its approximation error.  
		(a) High MOF case obtained with the damping coefficients $\alpha = 1 \times 10^{-3}$ and $\beta = 1 \times 10^{-2}$, resulting in MOF $\approx 1$ ($f =0.95\omega_5/(2\pi)$ = 32.8 Hz).  
		(b) Resonance case at the 5th mode under low MOF conditions, with $\alpha = 1 \times 10^{-4}$, $\beta = 1 \times 10^{-3}$ and $f =\omega_5/(2\pi) = 34.5 \text{ Hz}$.  
		For both (a) and (b), the panels show (Left) the full Gram matrix $\mathbf{G}_{full}$, (Middle) the approximate Gram matrix obtained by summing the contributions from four nearby modes (Modes 3–6), and (Right) the element-wise approximation error.
		}		
	\label{fig:highMOFres}
\end{figure}

Further analysis explores how MOF and resonance affect the structure of the Gram matrix. Figure~\ref{fig:highMOFres}(a) shows the result of increasing both $\alpha$ and $\beta$ by a factor of 10, yielding a high MOF of approximately 1. While the MOF increases with frequency and/or damping (see Eq.~\eqref{eq:MOF}), only the damping ratio was adjusted in this case, so that the effect on the overall structure of the Gram matrix can be maintained—compared to the low-MOF case shown in Figure~\ref{fig:gramDemo}(a). Figure~\ref{fig:highMOFres}(b) illustrates the case where the excitation frequency exactly matches the resonance frequency of the 5th mode. As the MOF is kept low (approximately 0.1) during this change, the comparison with Figure~\ref{fig:gramDemo}(a) highlights the effect of resonance.

As expected, the approximation error becomes large when the MOF is high (Figure~\ref{fig:highMOFres}(a), middle and right). Even under a low MOF, the resonance case results in extreme errors that far exceed the dynamic range of the color axis; an inset is included to display the error pattern (Figure~\ref{fig:highMOFres}(b), middle and right).

Although the modal approximation may break down, $\mathbf{G}_{full}$ still offers useful intuition for understanding the force reconstruction results presented later. Attention should be paid to the evolving of off-diagonal regions. In Figure~\ref{fig:highMOFres}(a, left), the off-diagonal components are suppressed, and the matrix becomes more diagonal-dominant. The behavior can be explained by revisiting the expression of the Gram matrix. As shown in Eq.~\eqref{eq:gram_expansion2}, the element $G_{ij}$ is governed by the balance between the self- and cross-terms, both involving complex-valued denominators. As the MOF increases, the imaginary parts of these denominators grow, reducing the magnitude of both terms. They tend to cancel each other out, especially when the spatial distance $|i - j|$ is large. The cancellation consequences to a Gram matrix that exhibits diagonal dominance. While a high MOF broadens the diagonal band—causing slight spreading around the true force location, it prevents the occurrence of spurious estimates.

Conversely, the resonance case shown in Figure~\ref{fig:highMOFres}(b, left) displays a tile-like pattern with strong off-diagonal components. The tile-like structure arises from the sole dominance of the fifth mode, whose outer-product $\boldsymbol{\phi}_5 \boldsymbol{\phi}_5^\mathsf{T}$ generates a uniform correlation pattern throughout the matrix. Then, spatially erroneous force estimates emerge, deviating far from the actual excitation location.

\begin{figure}[t]
	\centering
	\includegraphics[width=1\linewidth]{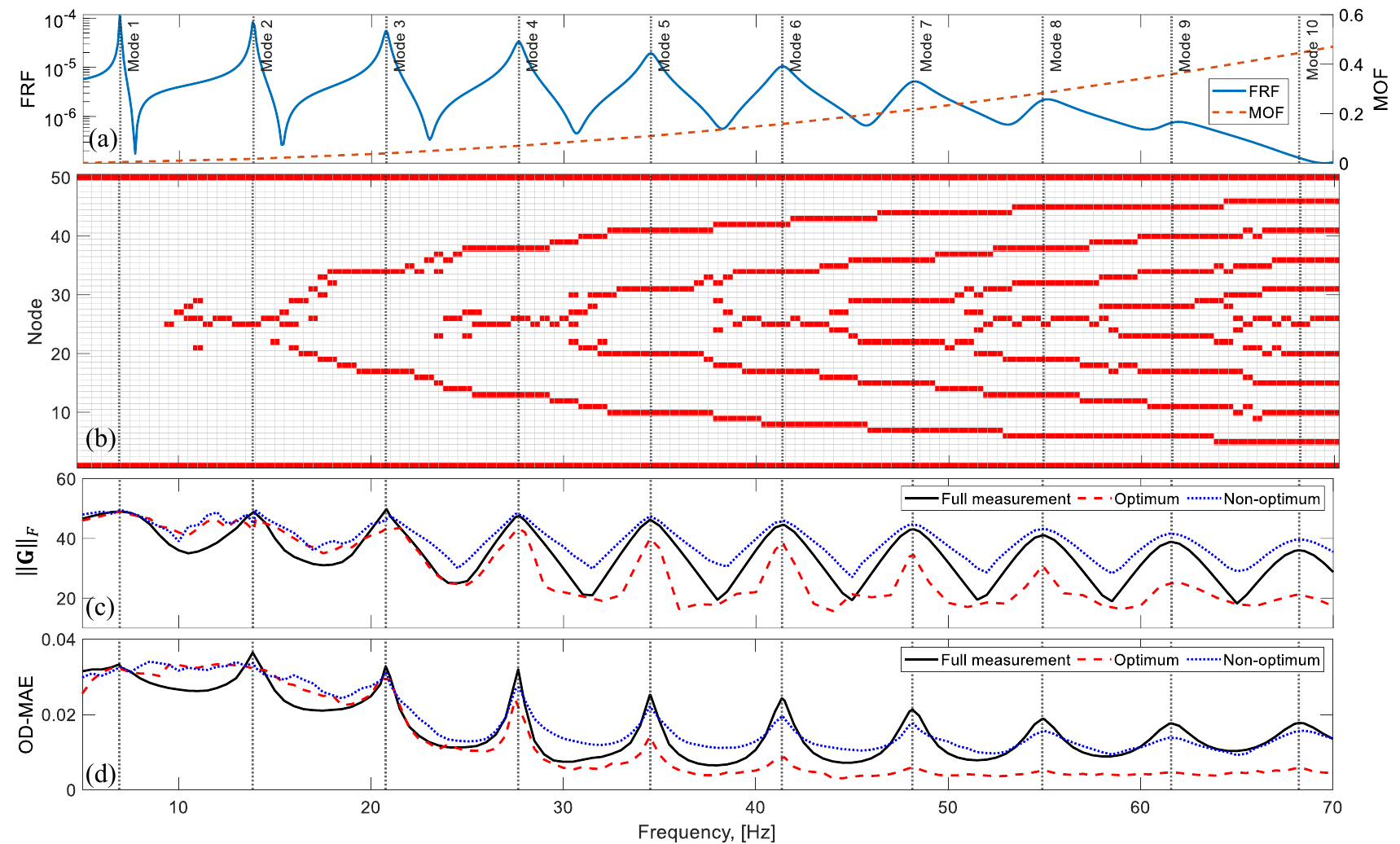}
\caption{Comparison of sensor placement strategies across a range of frequencies.  
	(a) Frequency response function (left axis) and modal overlap factor (right axis). Vertical lines indicate resonant frequencies.  
	(b) Activation map of optimal sensor locations. Red squares (\textcolor{red}{$\blacksquare$}) indicate the selected nodes.  
	 (c) Frobenius norm of the Gram matrix $\|\mathbf{G}(f)\|_F$ (Eq.~\eqref{eq:Gf}). (d) squared reconstruction errors (OD-MAE, Eq.~\eqref{eq:ODMAE})}

	\label{fig:overall}
\end{figure}

In summary, stable force recovery requires a Gram matrix with small off-diagonal components, a condition achieved under high MOF and off-resonance. Figure~\ref{fig:overall}(c, black solid line) presents the Frobenius norm of the full Gram matrix, defined as
\begin{equation}\label{eq:Gf}
	\| \mathbf{G} \|_F = \left( \sum_{i=1}^N \sum_{j=1}^N |G_{ij}|^2 \right)^{1/2},
\end{equation}
and computed over a range of frequencies. Consistent with the requirement for stable recovery, the norm exhibits periodic peaks at resonance frequencies and an overall decreasing trend due to the increasing MOF with frequency. While the results assume a full-measurement scenario in which all nodes are instrumented, it offers limited benefit over well-chosen sensor placements, as discussed in the next section.

\section{Optimal sensor placement}\label{sec:greedy}
\subsection{Greedy algorithm}
Placing sensors around the nodal regions of dominant mode shapes seems reasonable at or near resonance, but the locations become ambiguous when the target frequency lies between resonances. Sensor placement is guided by an optimization-based strategy, formulated as the following problem:
\begin{equation} \label{eq:SPproblem}
	\hat{S} = 
	\underset{S \subseteq \{1, 2,\, \ldots, N\},\; |S| = M}{\arg \min}
	\left\| \mathbf{G} - \text{diag}(\mathbf{G}) \right\|_F,
\end{equation}
where $S$ denotes the selected subset of rows from $\overline{\mathbf{H}}_{full}$. Sensor locations are chosen to minimize the Frobenius norm of the off-diagonal elements of the reduced Gram matrix. 

Choosing a subset is a combinatorial search, which does not fit the convex optimization. An alternative for obtaining locally optimal solutions in a reasonable time with less computational effort is to use the greedy algorithms~\cite[Sec. 3.2]{foucart2013}. The algorithm makes locally optimal decisions at each iteration, aiming to reach the global minimum at last, by adding the row that minimizes the cost to the subset $S$ at every step. Details are as follows:

\begin{enumerate}
	\item \textbf{Initialization}: Start with an empty set of selected rows, \( S = \emptyset \), and define the set of remaining rows as \( R = \{1, 2, \ldots, N\} \).
	
	\item \textbf{Iterative Selection}: At each iteration, 	
	 select the row \( i^* \in R \) that results in the smallest norm of \( \mathbf{G} - \text{diag}(\mathbf{G}) \) when added to \( S \):
	\begin{equation} \label{eq:iter1}
		i^* = \underset{i \in R}{\arg \min}\| \mathbf{G}_{S \cup \{i\}} - \text{diag}(\mathbf{G}_{S \cup \{i\}}) \|_F,
	\end{equation}
	where $ \mathbf{G}_{S \cup \{i\}} = \overline{\mathbf{H}}_{S \cup \{i\}}^\mathsf{H} \overline{\mathbf{H}}_{S \cup \{i\}}$, and $\overline{\mathbf{H}}_{S \cup \{i\}}$ is a sub-matrix of $\overline{\mathbf{H}}_{full}$ containing only the rows indexed by $S \cup \{i\}$. Then update the sets:
	\begin{equation}\label{eq:iter2}
		S \leftarrow S \cup \{i^*\}, \quad R \leftarrow R \setminus \{i^*\}.
	\end{equation}
	
	\item \textbf{Termination}: Repeat the selection process (Eqs.~(\ref{eq:iter1},\ref{eq:iter2})) until the size of $S$ satisfies $|S| = M$.
	
	\item \textbf{Output}: The final subset \( \hat{S} = S \) is the return value of the greedy algorithm.
\end{enumerate}
%
%\begin{algorithm}[t]
%	\caption{Greedy Sensor Selection Algorithm}
%	\begin{algorithmic}%	\label{alg:greedy}
%		\STATE \textbf{Initialization:} $S \gets \emptyset$; $R \gets \{1, 2, \ldots, N\}$
%		\WHILE{$|S| < M$}
%		\STATE Select $i^* = \arg \min\limits_{i \in R} \| \mathbf{G}_{S \cup \{i\}} - \text{diag}(\mathbf{G}_{S \cup \{i\}}) \|_2$
%		\STATE $S \gets S \cup \{i^*\}$; \quad $R \gets R \setminus \{i^*\}$
%		\ENDWHILE
%		\STATE \textbf{Output:} $\hat{S} \gets S$
%	\end{algorithmic}
%\end{algorithm}

\begin{algorithm}[htb]
	\caption{Greedy sensor selection algorithm}
	\label{alg:greedy}
	\SetKwInOut{Input}{Input}
	\SetKwInOut{Output}{Output}
	
	\Input{Full FRF matrix $\overline{\mathbf{H}}_{full} \in \mathbb{C}^{N \times N}$, desired number of sensors $M$}
	\Output{Selected sensor indices $\hat{S}$}

	$S \gets \emptyset,	R \gets \{1,2,\ldots,N\}$ 
	
	\While{$|S| < M$}{

		$i^* = \arg \min_{i \in R} \left\| \mathbf{G}_{S \cup \{i\}} - \text{diag}(\mathbf{G}_{S \cup \{i\}}) \right\|_F$
		
		%where $\mathbf{G}_{S \cup \{i\}} = \overline{\mathbf{H}}_{S \cup \{i\}}^\mathsf{H} \, \overline{\mathbf{H}}_{S \cup \{i\}}$\;
		
		$S \gets S \cup \{i^*\}$,\quad $R \gets R \setminus \{i^*\}$\;
	}
	\Return{$\hat{S} \gets S$}
\end{algorithm}
As summarized in Algorithm~\ref{alg:greedy}, the procedure begins with $M$, the number of sensors, which should be sufficient to cover all nodal regions of the dominant mode shape. Since the greedy algorithm does not guarantee a global optimum, choosing an $M$ well above the expected number of nodal points is recommended. It should also be noted that the algorithm is sequential rather than combinatorial; each iteration only tests the remaining candidates, thus increasing $M$ does not impose a computational burden.

\begin{figure} 
	\centering
	\includegraphics[width=1.0\linewidth]{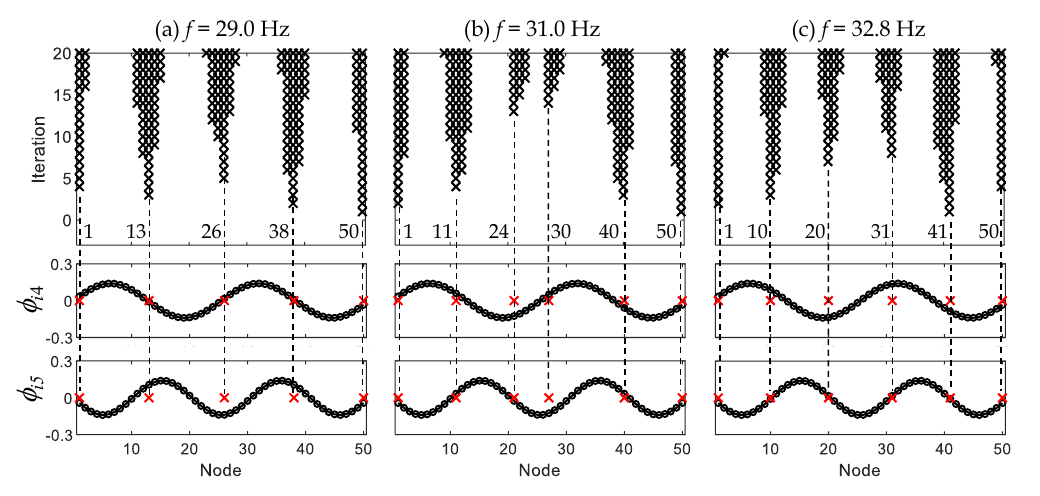}
	\caption{
		Greedy sensor selection results for three target frequencies: (a) 29 Hz ($= 1.05\omega_4/2\pi$), (b) 31 Hz (midpoint between $\omega_4$ and $\omega_5$), and (c) 32.8 Hz ($= 0.95\omega_5/2\pi$). 
		Top row: Iterative selection history over 20 steps. Middle and bottom rows: Mode shapes of the fourth and fifth modes, respectively, with final sensor locations (\textcolor{red}{$\boldsymbol{\times}$}) superimposed.	}
	
	\label{fig:fig5x}
\end{figure}

\subsection{Example}
Figure~\ref{fig:fig5x} illustrates how the greedy algorithm selects sensor locations for different target frequencies in the regular model (Figure~\ref{fig:spatial_model}).  The three columns correspond to distinct frequencies: (a) 29 Hz ($= 1.05\omega_4/2\pi$), (b) 31 Hz, and (c) 32.8 Hz ($= 0.95\omega_5/2\pi$). The top row shows the selection history over 20 iterations. During the iterations, selected nodes form clusters. 
Final sensors are assigned to the most frequently selected nodes in each cluster. Middle and bottom rows display the fourth and fifth mode shapes, respectively, with the final sensor positions superimposed (as indicated by \textcolor{red}{$\boldsymbol{\times}$}). In cases (a) and (c) where the target frequency lies near resonance, the sensors are located near the nodal regions of the dominant mode. Case (b) is of interest, as the target frequency is situated roughly halfway between the 4th and 5th resonances. The sensors are positioned along the nodal points of both modes.  This demonstrates the algorithm’s ability to accommodate the influence of contributing modes when no single mode is clearly dominant. 

\begin{figure}[t]
	\centering
	\includegraphics[width=0.8\linewidth]{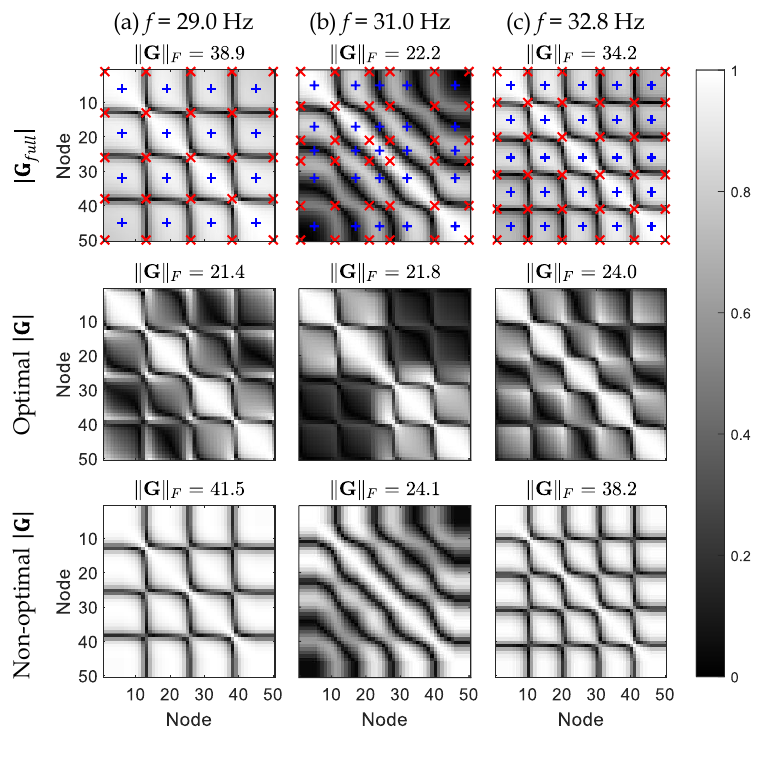}
	\caption{
		Comparison of Gram matrix structures for three target frequencies corresponding to those in Figure~\ref{fig:fig5x}. 	
		Top row: Full Gram matrix $\mathbf{G}_{full}$ with \textcolor{red}{$\boldsymbol{\times}$} indicating pairs of optimally selected sensor locations, and \textcolor{blue}{$\boldsymbol{+}$} denoting non-optimal (anti-nodal) pairs. 
		Middle (Bottom) row: Reduced Gram matrices using the optimal (non-optimal) sensor placements. 	
	}
	\label{fig:fig6x}
\end{figure}

To assess the effectiveness of the selected sensors, Figure~\ref{fig:fig6x} compares the Gram matrix for the target frequencies used in Figure~\ref{fig:fig5x}. The top row shows the full Gram matrix $\mathbf{G}_{full}$ at each frequency, with \textcolor{red}{$\boldsymbol{\times}$} indicating pairs of optimally selected sensor locations, and \textcolor{blue}{$\boldsymbol{+}$} non-optimal (anti-nodal) pairs. The second and third rows show the reduced Gram matrices constructed using the optimal (\textcolor{red}{$\boldsymbol{\times}$}) and the non-optimal  (\textcolor{blue}{$\boldsymbol{+}$}) locations, respectively. The optimal placements yield matrices with significantly lower off-diagonal energy. As annotated in each panel, the optimal placement yields the lowest Frobenius norm.

For a range of target frequencies, the optimal sensor locations are obtained by Algorithm~\ref{alg:greedy} and visualized in Figure~\ref{fig:overall}(b). At resonant frequencies, sensors are consistently placed near the nodal points of the mode shapes. Note that the end nodes are always selected due to the fixed boundaries. In non-resonant frequencies, the selected nodes tend to align with the nodal regions of nearby modes. Accordingly, the activation map exhibits a distinctive frequency-differentiated pattern.

The Frobenius norms of the Gram matrices for each sensor configuration are shown in Figure~\ref{fig:overall}(c). The red dashed and blue dotted lines represent the optimal and non-optimal (anti-nodal) sensor layouts, respectively. In the low-frequency region where the MOF is also low, the improvement achieved by the optimal configuration over the other two layouts is marginal. The low-frequency modes of the regular structure are characterized by long wavelengths and few nodal points, thus obtaining mutually incoherent measurements is difficult regardless of sensor placement. The columns of the resulting FRF matrix $\overline{\mathbf{H}}$ are highly correlated for all configurations, and even full measurement fails to ensure sufficient diversity.

In high frequencies with high MOF, the performance gap widens. The high-frequency modes provide more spatial diversity due to shorter wavelengths and an increased number of nodal regions. A few poorly placed or redundant sensors degrade the quality of the Gram matrix by introducing coherence. The proposed method is effective in selecting mutually incoherent measurements.

It should be noted that the above explanation holds for regular structures, whose mass, damping, and stiffness matrices have banded forms, as is the case for the structure under consideration. In irregular or geometrically complex systems, low-frequency modes can have spatially complex shapes and elevated MOF. In such cases, the proposed method proves effective even in the low-frequency regime (see Section~\ref{sec:sim_irregular}).

\section{Force reconstruction simulation} \label{sec:numerical}

\subsection{Regular structure example} \label{subsec:sim1}

\begin{figure}[t]
	\centering
	\includegraphics[width=0.95\linewidth]{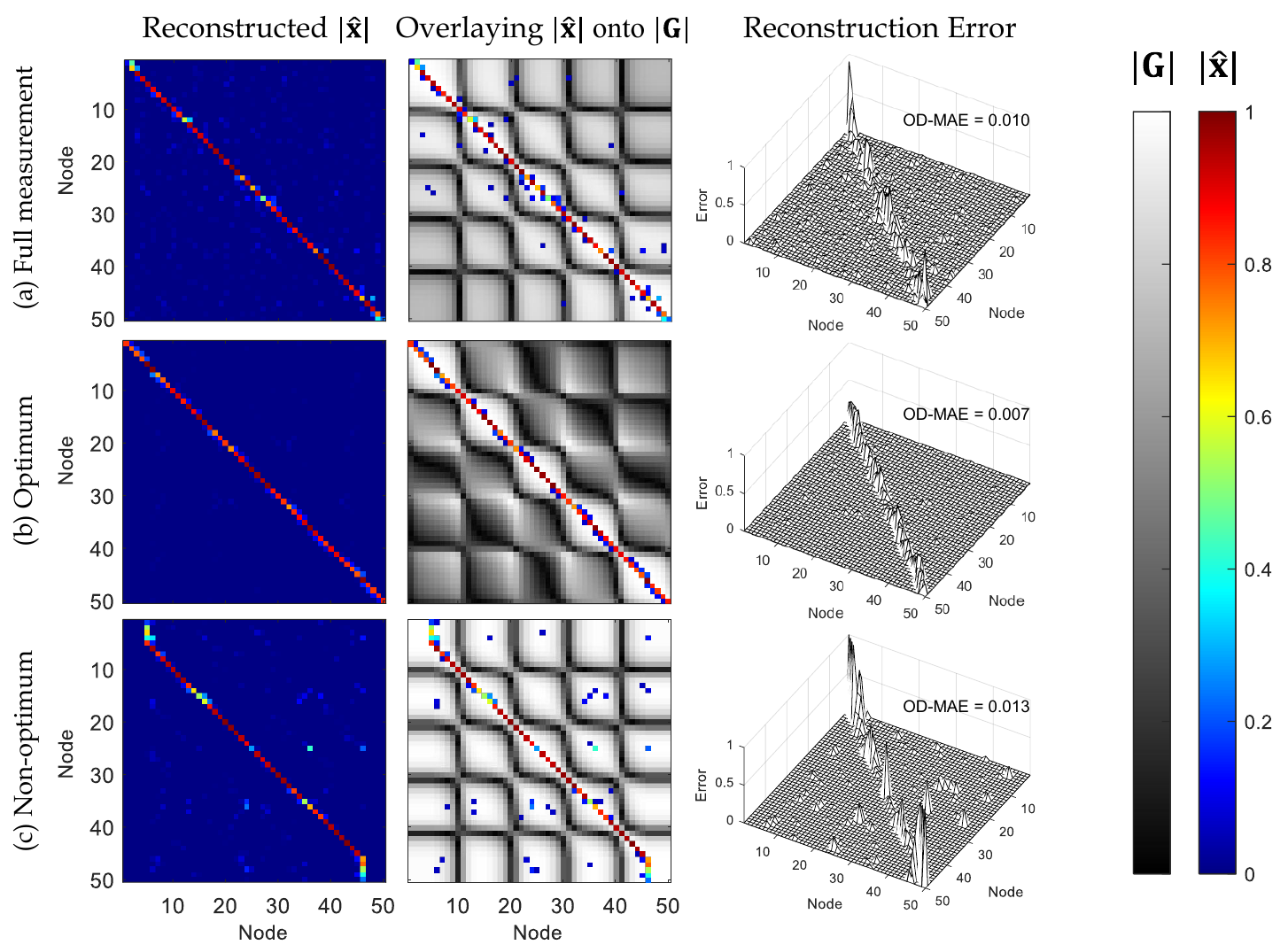}
\caption{
		Reconstruction performance for different sensor configurations—full measurement (top row), optimal placement (\textcolor{red}{$\boldsymbol{\times}$} markers in Figure~\ref{fig:fig5x}(c), middle row), and non-optimal placement (\textcolor{blue}{$\boldsymbol{+}$} markers in Figure~\ref{fig:fig5x}(c), bottom row)—at 32.8~Hz.
		Left column: estimated force maps ($\hat{\x}^{(i)}$), with the $i$-th row in each panel showing the result for a unit force applied at node $i$.  
		Middle column: reconstruction results overlaid on the Gram matrix, with values above a threshold of 0.01 shown.  
		Right column: reconstruction errors with the OD-MAE (Eq.~\eqref{eq:ODMAE}) values indicated .}
	\label{fig:verifySim0}
\end{figure}

Force reconstruction simulations are conducted using the regular structure in Figure~\ref{fig:spatial_model}. Given a sensor configuration, a unit-amplitude force is applied sequentially to the $i$-th node ($i = 1, \dots, N$), and the measurement vector is simulated according to Eq.~\eqref{eq:linear_model}. 
Additive white Gaussian noise is introduced with a signal-to-noise ratio defined as $\mathrm{SNR} = 20\log_{10}(\Vert \mathbf{H}\mathbf{x} \Vert_2 / \Vert \mathbf{n} \Vert_2)$, set to 20~dB. Reconstruction is performed by solving Eq.~\eqref{eq:lasso_normal} using CVX. After several trials, the regularization parameter is set to $\overline{\mu} = 0.1\big\Vert \overline{\mathbf{H}}^\mathsf{H}\overline{\mathbf{y}} \big\Vert_\infty$, following the L-curve method~\cite{kim2004}. This predetermined value is kept constant throughout the simulations to enable a fair comparison between sensor configurations. Varying the parameter by several tens of percent does not cause any notable change in the results below.

Corresponding to the case shown in Figure~\ref{fig:fig6x}(c), Figure~\ref{fig:verifySim0} presents reconstruction results at $f = 32.8$~Hz for three sensor configurations: full measurement (top row), optimal (nodal point based, middle row), and non-optimal (anti-nodal point based, bottom row). Each plot in the left column shows a 2D map constructed by stacking the denormalized force vectors $\hat{\x}^{(i)} = \Vert \y \Vert_2 \mathbf{F} \hat{\overline{\x}}^{(i)}$ row-wise, where the $i$-th row represents the $i$-th force input. An ideal reconstruction yields an identity matrix $\mathbf{I}_N$. The right column shows the reconstruction errors, i.e., the deviation from $\mathbf{I}_N$.

\begin{figure}[t]
	\centering
	\includegraphics[width=1\linewidth]{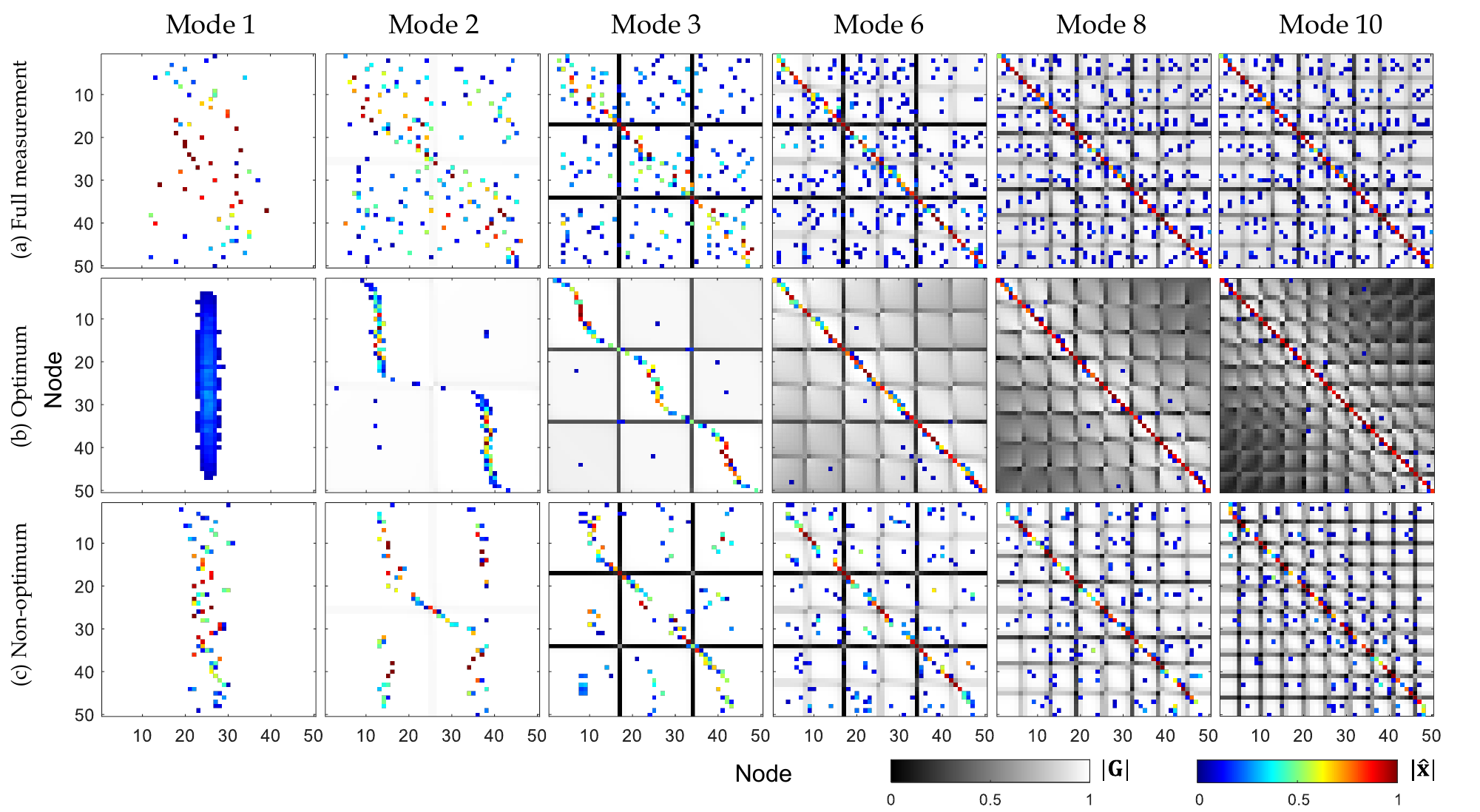}
	\caption{Comparison of reconstruction results at resonance frequencies for three sensor configurations: (a) full measurement (top row), (b) optimal placement (middle row), and (c) non-optimal (anti-nodal) placement (bottom row). In each plot, the denormalized force vectors $|\hat{\mathbf{x}}|$ (color scale) are overlaid on the Gram matrix (grayscale).}
	\label{fig:fig8x}
\end{figure}

Most errors are concentrated near the diagonal, indicating underestimation of true force magnitudes due to the soft-thresholding effect inherent to the $\ell_1$-norm penalty~\cite{wright2009,lee2025}. While this can be mitigated by employing non-convex penalties~\cite{qiao2019,aucejo2016}, such approaches introduce non-convexity of the cost function and may hinder global convergence. The focus is on suppressing off-diagonal entries across different sensor setups. Both full and non-optimal placements produce spurious estimates in regions where no actual force is present. Those ghosts are likely to occur at positions with high off-diagonal components of the Gram matrix. The agreement is visualized by overlaying the reconstruction maps on the Gram matrices (middle column), showing only values above a threshold of 0.01 for clarity. 

Reconstruction accuracy is quantified using the off-diagonal mean absolute error (OD-MAE), defined as:
\begin{equation}\label{eq:ODMAE}
	\mathrm{OD\text{-}MAE} = \frac{1}{N(N-1)} \sum_{\substack{i,j=1 \\ i \neq j}}^{N} \left| \hat{x}_{ij} \right|,
\end{equation}
which measures the average estimated force at off-target locations. As depicted in the error plot, the optimal configuration gives the lowest OD-MAE among the considered setups.

Figure~\ref{fig:overall}(d) compares the OD-MAE values over the frequency range for three sensor layouts. Similar to the behavior of the Frobenius norm $\|\mathbf{G}\|_F$ shown in Figure~\ref{fig:overall}(c), all three curves  exhibits periodic peaks at resonance frequencies and an overall decreasing tendency, attributed to the increasing modal overlap factor (MOF) with frequency. The optimal configuration achieves substantial error reduction in high-MOF regions, even at resonant frequencies. 

To further assess the reconstruction performance at resonance frequencies, detailed results for each mode are presented in Figure~\ref{fig:fig8x}. As noted earlier, resonance conditions produce a tile-like Gram matrix becoming more refined at higher frequencies due to shorter wavelengths. Pronounced off-diagonal components degrade reconstruction accuracy, particularly when sensor placement is suboptimal. For modes with high MOF (Modes 6–10), the optimal sensor configuration (middle row) effectively suppresses off-diagonal components of the Gram matrix, achieving accurate and focused force estimates.

For low-frequency modes (Modes 1–3), all sensor configurations exhibit poor reconstruction accuracy. But, the failure patterns differ depending on the sensor counts. In the full configuration where the number of sensors matches the number of unknowns, the optimization problem in Eq.~\eqref{eq:lasso_normal} reduces to solving a square linear system. The resulting solution becomes more sensitive to measurement noise than to the coherence of the FRF matrix, causing widely scattered estimates. For a limited number of sensors (either optimal or non-optimal), the reconstruction is more influenced by the coherence structure, and the estimates concentrate in regions of high Gram matrix correlation. These results reaffirm the challenge of force reconstruction in low-MOF regime.

\subsection{Irregular structure example} \label{sec:sim_irregular}

To demonstrate the generality of the proposed method beyond simple models, its performance is evaluated using an irregular system with randomized yet physically plausible mass and stiffness matrices. The number of dof is maintained at $N = 50$. The mass matrix is defined as $\mathbf{M} = \mathbf{R}^\mathsf{T} \mathbf{R}$, where $\mathbf{R} \in \mathbb{R}^{N \times N}$ is a Gaussian random matrix, ensuring that $\mathbf{M}$ is symmetric and positive-definite. The stiffness matrix is given by $\mathbf{K} = \mathbf{Q} \boldsymbol{\Lambda} \mathbf{Q}^\mathsf{T}$,  
where $\mathbf{Q}$ is a random orthonormal matrix obtained via the QR decomposition of a Gaussian random matrix,  
and $\boldsymbol{\Lambda}$ is a diagonal matrix with positive entries increasing linearly from $10^5$ to $10^6$. This construction guarantees a symmetric positive-definite $\mathbf{K}$ while inducing nonuniform mode shapes due to the randomized eigenvectors in $\mathbf{Q}$. The damping matrix $\mathbf{C}$ is constructed as a symmetric random matrix to represent general viscous damping without assuming proportionality.  
While the exact damping mechanism is not specified, the  $\mathbf{C}$ matrix is scaled such that the resulting modal damping ratios $\zeta_r$ fall within the range of $[0.01, 0.1]$.

\begin{figure}[t]
	\centering
	\includegraphics[width=0.7\linewidth]{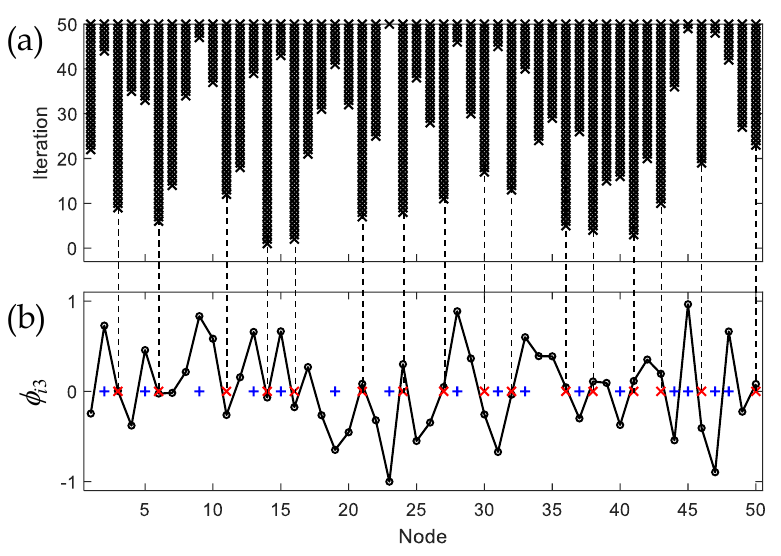}
	\caption{Sensor selection results at the target frequency of 6.4~Hz (near the third mode at 6.9~Hz) for the irregular structural model.  (a) Sensor selection history over 50 iterations using Algorithm~\ref{alg:greedy}.  (b) The third mode shape at 6.9~Hz.  The optimal sensor locations (\textcolor{red}{$\boldsymbol{\times}$}) align with the nodal regions, while the non-optimal (anti-nodal) configuration is indicated by \textcolor{blue}{$\boldsymbol{+}$}. Both configurations use 16 sensors.}
	\label{fig:fig9x}
\end{figure}

A target frequency of 6.4~Hz is chosen near the third resonance at 6.9~Hz. Several neighboring modes exist within a $\pm$3~Hz window around this frequency, resulting in a high MOF ($\cong 1$). Despite being in the low-frequency range, the dominant (third) mode exhibits a complex shape due to structural irregularity, as shown in Figure~\ref{fig:fig9x}(b).  So, the sensor selection history obtained using Algorithm~\ref{alg:greedy} reveals multiple clusters, implying the need for a larger number of sensors. The optimal configuration selects 16 sensors (marked by \textcolor{red}{$\boldsymbol{\times}$}), aligning with the nodal regions. Other frequently selected nodes, though not included in the final set, could be viable candidates. Their inclusion has little impact on the reconstruction results presented below.

The subsequent validation procedure follows the same setup as described in Section~\ref{subsec:sim1}.  Three sensor configurations are tested: full measurement (50 sensors), optimal (\textcolor{red}{$\boldsymbol{\times}$}), and non-optimal (anti-nodal, \textcolor{blue}{$\boldsymbol{+}$}) placements, both using 16 sensors. Each unit-amplitude force is applied sequentially to all nodes, with the output signal simulated via Eq.~\eqref{eq:linear_model} at an SNR of 20~dB. The regularization parameter for solving Eq.~\eqref{eq:lasso_normal} is fixed at $\overline{\mu} = 0.1 \Vert \overline{\mathbf{H}}^\mathsf{H} \overline{\mathbf{y}} \Vert_\infty$.

\begin{figure}[t]
	\centering
	\includegraphics[width=1\linewidth]{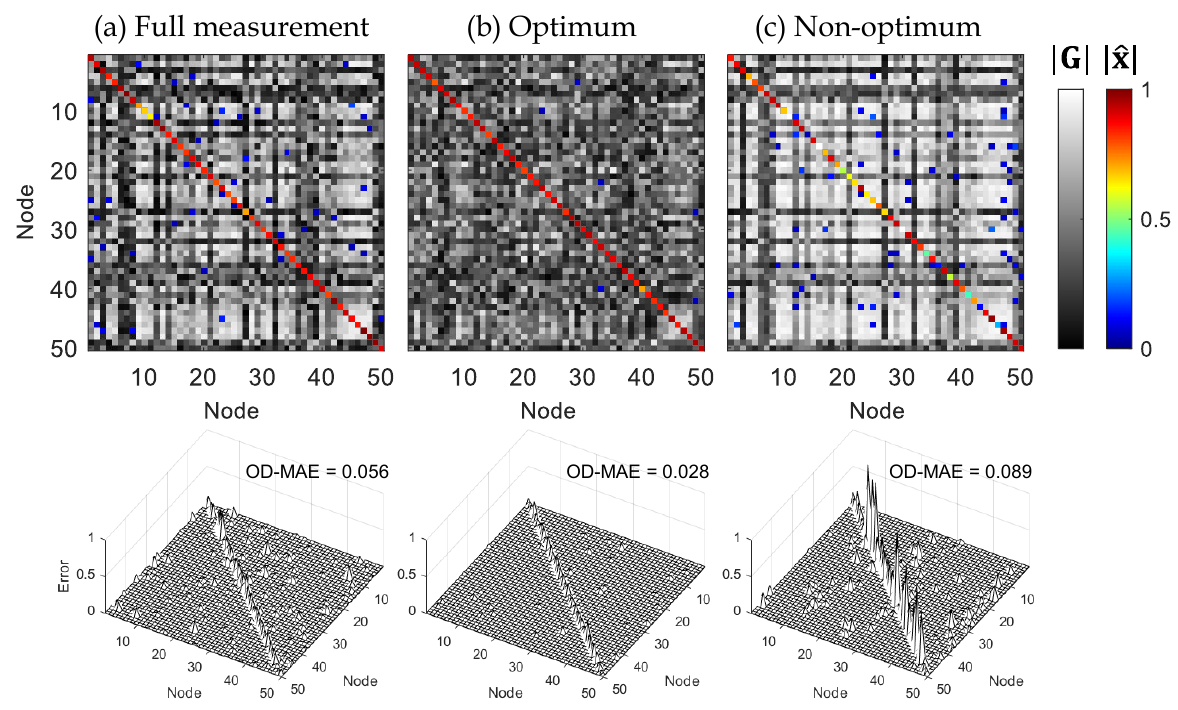}
	\caption{
		Reconstruction results for three sensor configurations on the irregular model:  
		(a) full measurement, (b) optimal placement, and (c) non-optimal (anti-nodal) placement, all evaluated at 6.4~Hz.  
		Top row: estimated force distributions overlaid on the Gram matrix, with values above a threshold of 0.01 shown.  
		Bottom row: Reconstruction errors.  
	}
	
	\label{fig:fig10x}
\end{figure}

Figure~\ref{fig:fig10x} compares the performance of three sensor configurations. The Gram matrices in the top row reveal clear differences in sensing quality. The optimal configuration yields the lowest off-diagonal energy. The full and non-optimal configurations exhibit stronger off-diagonal components in $\mathbf{G}$, suggesting greater correlation among sensors. These characteristics are manifested in the overlaid reconstructions and the error plot below. The optimal configuration achieves the best performance (OD-MAE = 0.028), with diagonal errors arising from the soft-thresholding effect of the $\ell_1$-norm penalty~\cite{wright2009,lee2025}. The full configuration shows moderate accuracy (OD-MAE = 0.056) despite its dense sensor coverage. The non-optimal configuration performs the worst (OD-MAE = 0.089), producing spurious estimates not only far from the diagonal but also with significant errors along the diagonal. Optimal sensor placement proves effective for reconstruction accuracy. 

An interesting observation is that, even at low frequencies, the irregular structure exhibits relatively high modal overlap due to the proximity of modes induced by randomness in mass and stiffness distributions. The increased MOF enhances the effectiveness of the proposed method, as it benefits from richer spatial variation in the mode shapes. Hence, the nodal-point-based sensor placement yields accurate reconstruction, in contrast to the performance degradation seen in the low-frequency region of the regular structure.

\section{Experiment} \label{sec:exp}

\begin{figure}[t]
	\centering
	\includegraphics[width=.9\linewidth]{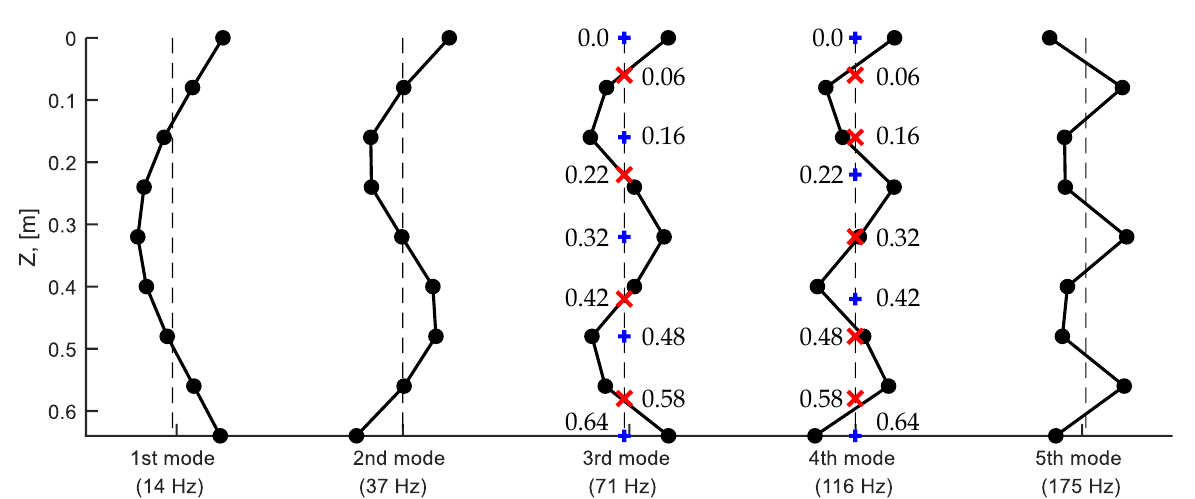}
	\caption{ Experimental mode shapes of the test beam. The third and fourth modes, resonating near 65 and 110~Hz, respectively, were used to select the sensor configurations. Nodal-based optimal positions are marked with \textcolor{red}{$\boldsymbol{\times}$}, anti-nodal (non-optimal) positions with \textcolor{blue}{$\boldsymbol{+}$}, and the full sensor configuration includes all of these locations.}
	  \label{fig:expmodeshape}
\end{figure}

Experimental validation was conducted for a steel ruler beam (64 $\times$ 3 $\times$ 0.01 cm, length $\times$ width $\times$ thickness) under free boundary conditions, implemented by attaching one end to a soft spring and suspending the beam freely. Figure~\ref{fig:expmodeshape} shows the first five flexible mode shapes, obtained through an experimental modal analysis using an impact hammer. The beam shares similar modal characteristics with the regular model discussed in Section~\ref{subsec:sim1}, allowing for similar interpretation of nodal regions and Gram matrix behavior. The validation considers two target frequencies: $f = 65$ and $f = 110$ Hz, both close to the third and fourth modes, respectively.  As frequency increases, more sensors are needed to capture the complex mode shapes. Those frequencies were selected in consideration of the channel capacity of the employed data acquisition system (Model: Siemens SCADAS).

To simplify the experimental procedure, nine accelerometers (Model: PCB 352C66) were installed at $Z = 0, 6, 16, 22, 32, 42, 48, 58$, and $64$ cm, while only the excitation frequency was varied. For the $f = 65$ Hz case, the sensor positions at $Z = 6, 22, 42$, and $58$ cm serve as the optimal configuration (\textcolor{red}{$\boldsymbol{\times}$}), whereas the remaining five positions (\textcolor{blue}{$\boldsymbol{+}$}) represent an anti-nodal (non-optimal) layout. Similarly, for the $f = 110$ Hz case, the optimal configuration comprises sensors at $Z = 6, 16, 32, 48$, and $58$ cm. While Algorithm~\ref{alg:greedy} can be used to determine the optimal placement more rigorously, in these particular cases, the modes are well-separated and the nodal regions are distinct, allowing near-optimal locations to be selected without explicit optimization.

\begin{figure}[t]
	\centering
	\includegraphics[width=0.4\linewidth]{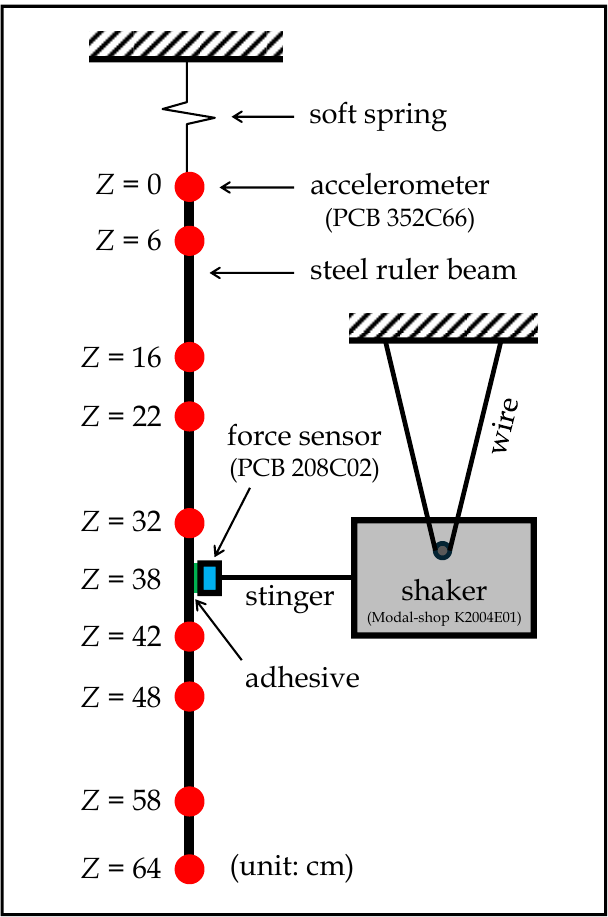}
	\caption{Schematic of the experimental setup. The beam is suspended vertically using a soft spring. Nine accelerometers (Model: PCB 352C66) are placed along the beam at designated locations. A force sensor (Model: PCB 208C02) is attached using adhesive at $Z = 38$~cm, in line via a stinger with a mini-shaker (Model: Modal-shop K2004E01).
	}  \label{fig:expsetup}
\end{figure}

Potential force locations were distributed at 2 cm intervals along the beam span, resulting in a total of 33 equally spaced nodes ($N = 33$). For nine acceleration measurements $(M=9)$, the FRF matrix $\mathbf{H}(\omega)$ has dimensions $9 \times 33$. Each individual transfer function $h_{mn}(\omega)$ was measured through an impact test, covering frequencies up to 512~Hz and utilizing the $H_1$ estimator~\cite{bendat2010}.

After the FRF measurements, the beam was excited at node 20 ($Z = 38$ cm) using a mini-shaker (Model: Modal-shop K2004E01) connected via a thin, flexible stinger. A force sensor (Model: PCB 208C02) was mounted in line with the shaker to monitor the input force. Suspended with a wire, the shaker remained stationary during excitation due to its large inertia relative to the lightweight beam. For a verification of the test setup, a tri-axial accelerometer was temporarily installed at the excitation point prior to the main experiments. The measured lateral accelerations were found to be less than 2\% of the axial component, confirming that off-axis excitation was negligible. A continuous sinusoidal signal at the target frequency (either 65 or 110 Hz) was generated using a function generator, fed to the shaker. The steady-state responses were recorded at all nine sensor locations along with the force input. The overall experimental setup is illustrated in Figure~\ref{fig:expsetup}.

\begin{figure}[t]  % 'p' 옵션은 새 페이지에 강제로 배치
	\centering
	\includegraphics[height=0.25\textheight]{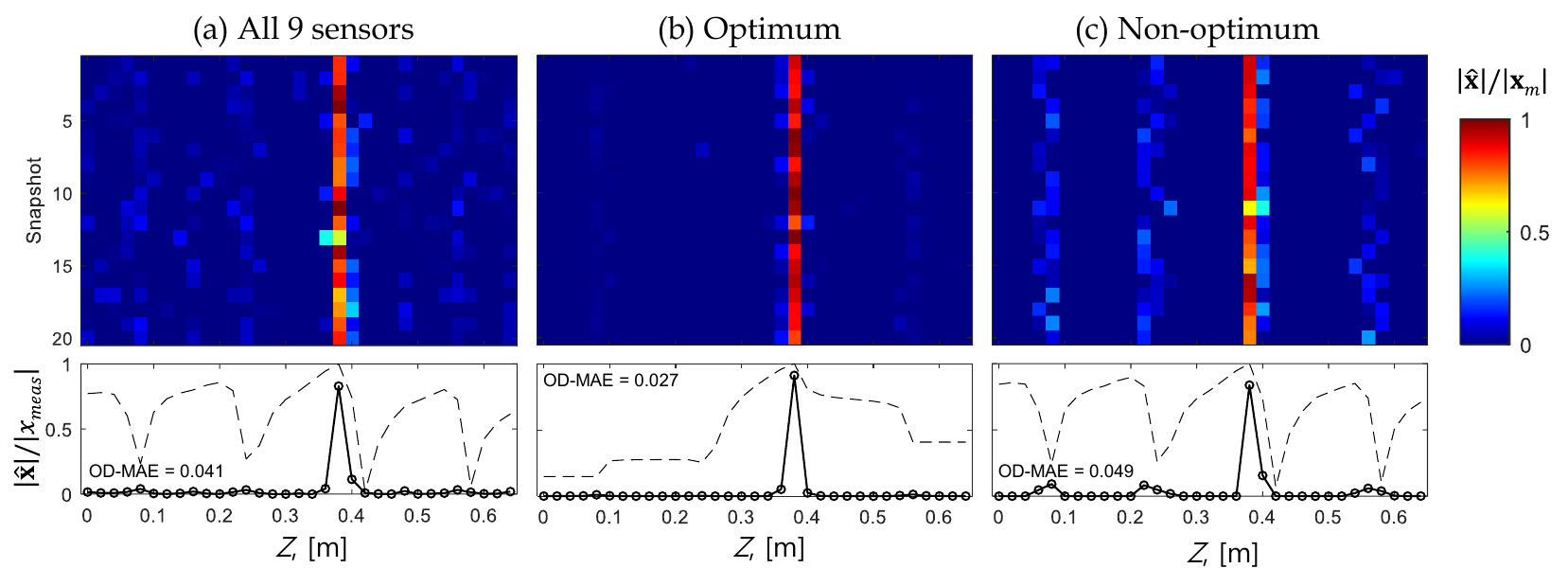}
	%	\caption{(Top) Force reconstruction results normalized by the actual force measurements $\x_m$ across snapshots (Bottom) Snapshot averaging (line with circle marker) and the Gram matrix (dashed line) with respect to the excitation point at $Z$ = 38 [cm] (Node 20). (a) Nodal-point scheme, (b) Anti-nodal point scheme.}
	\caption{Reconstruction results at 65~Hz. (Top) Snapshot-wise force estimates normalized by the measured input force $x_{\text{meas}}$. (Bottom) Snapshot-averaged estimate (solid line with circle markers) and Gram matrix profile (dashed line) with respect to the excitation point at $Z = 38$~cm. (a) All 9 sensors, (b) Optimal configuration, (c) Non-optimal configuration.}
	
	\label{fig:exp65Hz}
	
	\vspace{0.02\textheight} 
	
	\includegraphics[height=0.25\textheight]{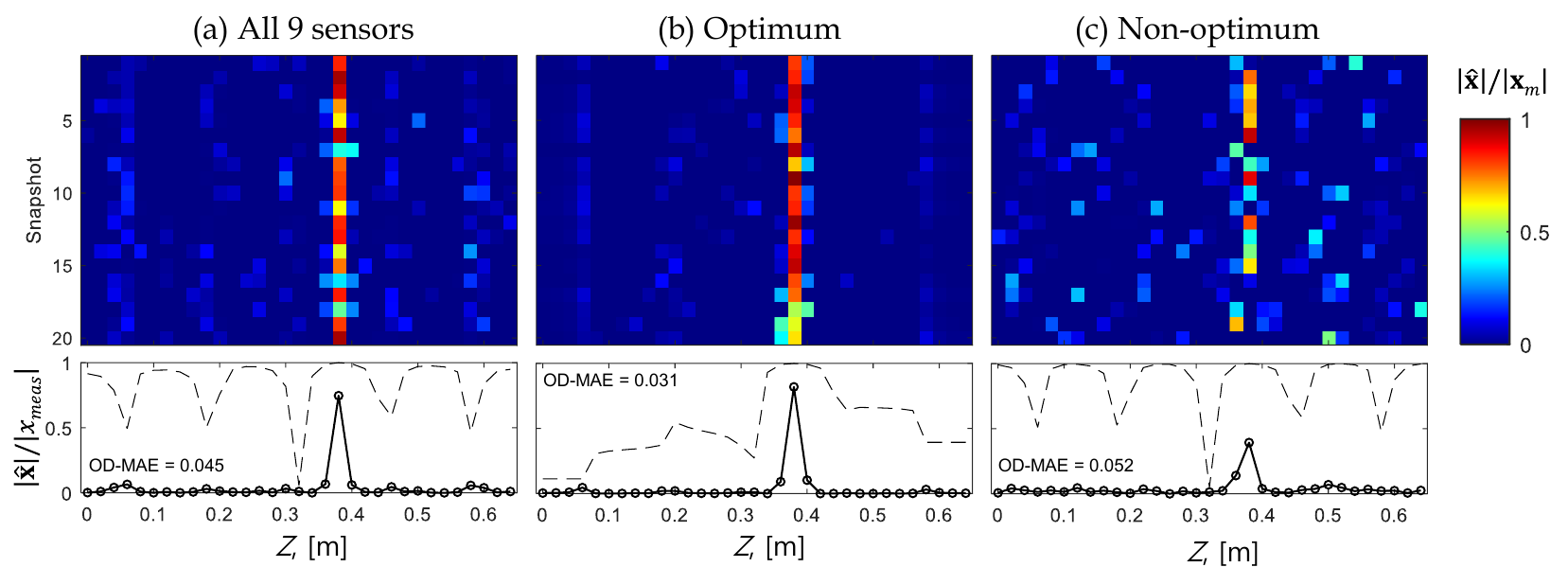}
	\caption{Reconstruction results at 110~Hz, presented in the same format as Figure~\ref{fig:exp65Hz}.}

	\label{fig:exp110Hz}
\end{figure}

For the acquired signal, a 1024-point fast Fourier transform was performed to obtain the measurement vector $\y$ at the target frequency. Following the simulation procedure, three sensor configurations were tested: all 9 sensors (mimicking the full measurement), an optimal (nodal) layout, and a non-optimal (anti-nodal) layout. With $\overline{\mu} = 0.08\Vert \overline{\mathbf{H}}^\mathsf{H}\overline{\y}\Vert_\infty$, Eq.~\eqref{eq:lasso_normal} was solved to estimate the force distribution. Figures~\ref{fig:exp65Hz} and~\ref{fig:exp110Hz} present the reconstruction results at 65~Hz and 110~Hz, respectively. In each figure, the top row shows snapshot-wise reconstructed force distributions normalized by the measured input force $x_{meas}$, and the bottom row compares the averaged estimates and the Gram matrix profiles with respect to the excitation node. 

The ideal reconstruction should be unity at the excitation location and zero elsewhere. The error characteristics are consistent with those observed in the simulations: spurious estimates arise near regions of high coherence, while the amplitude at the force location is underestimated due to the soft-thresholding effect of the $\ell_1$-norm penalty. At 65Hz (Figure~\ref{fig:exp65Hz}), the optimal configuration achieves sharp localization at the excitation point and minimal spurious components, whereas both the all-9 sensor measurement and the non-optimal layout yield dispersed estimates with off-target peaks. At 110Hz (Figure~\ref{fig:exp110Hz}), similar tendencies are observed. The accuracy declines slightly as the dominant (4th) mode exhibits greater spatial variation, making the reconstruction sensitive to sensor placement. Nevertheless, the optimal configuration still provides the best performance among the three cases, as confirmed by the OD-MAE values.

\section{Conclusion} \label{sec:conclusion}
This study has presented a systematic sensor placement method for sparse force reconstruction. The core idea lies in leveraging the Gram matrix, quantifying the coherence between the basis vectors of FRF matrix.  By employing a greedy algorithm to minimize the off-diagonal elements of the Gram matrix, the method  identifies sensor locations at or near the nodal points of the dominant mode shape associated with the target frequency. From a physical perspective, these locations correspond to regions of low spatial correlation, promoting independence among the measurements. Beyond its theoretical contribution, the method is applicable to scenarios including dynamic load identification, monitoring of operational forces in service structures, and noise/vibration control in complex engineering systems. 

While the proposed method demonstrates promising performance for single-frequency applications or narrowband scenarios, its extension to broadband or multi-frequency cases is restricted. Since the Gram matrix is inherently frequency-dependent, the optimal sensor configuration varies significantly across frequencies (Figure~\ref{fig:overall}b). A fixed layout optimized for a specific frequency does not guarantee robust performance over a wide frequency band. Possible alternatives include composite sensor sets that cover multiple nodal regions across frequencies, or multi-frequency optimization strategies that jointly consider several target frequencies.

In addition, the greedy algorithm adopted in this study is not the only viable option for sensor placement. Various alternatives can be considered to enhance incoherence in the FRF matrix. For example, one may adapt classical observability-based methods, such as the Effective Independence (EI) index~\cite{kammer1991}, not in its usual maximization form, which favors anti-nodal regions, but in a reversed formulation aimed at minimizing correlation. Such inverted criteria may yield sensor layouts similar to those obtained in this study. Although these alternative strategies were not implemented here, their comparison with the present Gram-matrix-based approach would be interesting.

On a practical note, this work did not examine certain implementation issues in detail, such as positional errors in sensor placement or the systematic tuning of the regularization parameter. These aspects, along with the generalization of the framework for broadband applications and its benchmarking against alternative placement methodologies, will be addressed in upcoming researches.

\section*{CRediT authorship contribution statement}
Jeunghoon Lee is the sole author of this paper.

\section*{Declaration of competing interest}
The authors have no conflicts to disclose.

\section*{Acknowledgments}
This work was supported by the Korea Institute of Marine Science and Technology Promotion (KIMST) Grant 20210500, and by the Ministry of Trade, Industry \& Energy (MOTIE, Korea) Grant RS-2025-24533624.

\section*{Data availability}
The data supporting the findings of this study are available from the author upon request.

\bibliographystyle{unsrt} %e.g., plain, alpha, IEEEtran

\bibliography{mssp2025bib}

\end{document}